\documentclass[hyper]{JHEP} 

\usepackage{epsfig}

\newcommand\fverb{\setbox\pippobox=\hbox\bgroup\verb}

\newcommand\fverbdo{\egroup\medskip\noindent%

            \fbox{\unhbox\pippobox}\ }

\newcommand\fverbit{\egroup\item[\fbox{\unhbox\pippobox}]}

\newbox\pippobox

\title{Note About Integrability of Non-Relativistic String}
\author{J. Kluso\v{n}\\
    Department of
    Theoretical Physics and Astrophysics\\
    Faculty of Science, Masaryk University\\
    Kotl\'{a}\v{r}sk\'{a} 2, 611 37, Brno\\
    Czech Republic\\
    E-mail: \email{klu@physics.muni.cz}} \preprint{}

\abstract{We define non-relativistic string in the background with
non-trivial metric. Then we analyze its properties when we focus on
integrability of this string in $AdS_5\times S^5$ background.}

\def\tr{\mathrm{Tr}}
\def\bT{\mathbf{T}}

\def\hE{\hat{E}}

\def\tx{\tilde{x}}

\def\pb  #1{\left\{#1\right\}}

\newcommand{\bK}{\mathbf{K}}

\def\tmH{\tilde{\mH}}

\def\tit{\tilde{t}}
\def\bM{\mathbf{M}}

\newcommand{\mH}{\mathcal{H}}
\newcommand{\mG}{\mathcal{G}}

\newcommand{\mL}{\mathcal{L}}

\newcommand{\hg}{\hat{g}}

\newcommand{\tlambda}{\tilde{\lambda}}

\newcommand{\hJ}{\hat{J}}

\newcommand{\tphi}{\tilde{\phi}}

\newcommand{\mY}{\mathcal{Y}}
\newcommand{\mZ}{\mathcal{Z}}

\def\ttau{\tilde{\tau}}
\def\ba{\mathbf{a}}
\begin{document}
\section{Introduction}
Recently systems with non-relativistic symmetries were  intensively studied from different points of view and in different situations. For example,  condensed matter systems are analyzed
\footnote{For recent review, see \cite{Hartnoll:2016apf}.}
 using holographic mapping since
it is well known that holography is useful tool for the study of relativistic
strongly coupled field theory using Einstein classical
gravity in the bulk \cite{Maldacena:1997re,Gubser:1998bc,Witten:1998qj}. Non-relativistic symmetry plays also fundamental role in the  famous P. Ho\v{r}ava's proposal
\cite{Horava:2009uw} of renormalizable Quantum Gravity,  for recent review, see \cite{Wang:2017brl}. Non-relativistic gravities were also analyzed recently in
    \cite{Bergshoeff:2017btm,Bergshoeff:2016lwr,Hartong:2015xda,
        Bergshoeff:2015uaa,Andringa:2010it,Bergshoeff:2015ija,
        Bergshoeff:2014uea}.

Non-relativistic theory can be also defined  when we perform specific limiting  procedure on the level of relativistic actions for fundamental strings and Dp-branes \cite{Gomis:2000bd,Danielsson:2000gi,Gomis:2004pw,Gomis:2005pg,Kluson:2006xi,Batlle:2016iel,Gomis:2016zur,Kluson:2017vwp}
\footnote{
Another very interesting limit that can be performed on the world-volume
of extended objects in string theory is Carroll limit, see for example
\cite{Bergshoeff:2014jla,Bergshoeff:2015wma,Cardona:2016ytk,Kluson:2017fam}.}.
Further, non-relativistic limits can  lead to new solvable sectors of string theory. Famous example is the analysis performed in  \cite{Gomis:2005pg} that leads to a new solvable sector of superstring theory on $AdS_5\times S^5$. Very recently
interesting proposal of  alternative definition of non-relativistic string was introduced in \cite{Harmark:2017rpg} and its relation with the dual sectors of CFT was discussed.

The goal of this paper is to analyze non-relativistic limit of fundamental
string in the background with non-trivial metric, as was defined recently in
\cite{Batlle:2016iel}. We begin with the Nambu-Gotto action for fundamental string in the background with non-trivial metric and perform non-relativistic limiting procedure as in \cite{Batlle:2016iel}. However due to the fact that the target space-time metric is not flat   we find that there is a non-trivial divergent term in the action and we argue that it can be canceled when we introduce
background NSNS two form that electrically couples to Nambu-Gotto string. Then we find that the limiting procedure is well defined and we obtain an action for non-relativistic  string in general background. Using this form of the action we analyze whether  it is possible to impose uniform gauge fixing for this string with analogy with the relativistic case \cite{Arutyunov:2004yx,Arutyunov:2006gs}, for review, see \cite{Arutyunov:2009ga}.
In order to do this we have to find Hamiltonian form of non-relativistic string
in general background and then to fix two first class constraints. After then we find the uniform gauge fixed Hamiltonian for non-relativistic string
and study its properties.

As the next step we focus on the integrability property of the non-relativistic string. In the first case we study the situation where  metric in the directions where  non-relativistic limit was performed is flat while the metric in the  transverse space corresponds to some group manifold. In this case we very easily find that the non-relativistic string is integrable. As the second case we consider non-relativistic string in $AdS_5\times S^5$ where we perform non-relativistic limit along time direction in $AdS_5$ and one angular direction in $S^5$. We determine the form of the background NSNS two form that is needed in order to be this procedure well defined. Then we use the factorization property of the coset representative \cite{Arutyunov:2004yx}
to find non-relativistic string on $AdS_5\times S^5$ background. We find Lax connection and we argue that it is flat when all world-sheet modes obey the equations of motion. We also determine Poisson brackets between spatial components of Lax connection and we argue that it takes the standard form that implies an existence of  infinite number of conserved charges that are in involution.

Let us outline our results and suggest possible open problems and directions for further study. We find an action for non-relativistic string in general background and we analyze its properties. Then we study integrability of this string and we argue that it can be integrable and find corresponding Lax connection. However it is important to stress one open  problem in this construction  that is related to  an existence of specific background NSNS two form which is needed in order to
cancel divergence that arises in the non-relativistic procedure. The problem is that it is not clear whether this background field  solves supergravity equations of motion when it depends on the transverse coordinates. In case when this two form is constant we find that the equations of motion are obeyed but the general case deserves further study. It would be also nice to see whether non-relativistic string has giant magnon like solutions and it would be nice to find corresponding dispersion relations
\cite{Hofman:2006xt}. We hope to return to these problems in future.

The structure of this paper is as follows. In the next section (\ref{second}) we introduce non-relativistic string in the general background and study its properties. Then in section (\ref{third}) we find canonical form of this action and determine its constraint structure. In section (\ref{fourth}) we perform uniform gauge fixing of this string and determine corresponding Hamiltonian. In section (\ref{fifth}) we analyze integrability of non-relativistic string.

\section{Non-Relativistic String in Non-Trivial Background}\label{second}
Let us consider Nambu-Gotto action for fundamental string
\begin{equation}\label{NGaction}
S=-\ttau_F\int d\sigma d\tau \sqrt{-\det g_{\alpha\beta}} \ , \quad
g_{\alpha\beta}=G_{MN}\partial_\alpha \tx^M\partial_\beta
\tx^N+G_{\mu\nu}\partial_\alpha \tx^\mu\partial_\beta \tx^\nu \ ,
\end{equation}
where  $G_{MN}$ is the metric on the
transverse space with coordinates $\tx^M, \  M,N=2,\dots,9$ that depend on $\tx^M$ but not
on $\tx^\mu$.  We presume that the metric $G_{\mu\nu}$ does not depend on
$\tx^0\equiv \tilde{t}$ and $\tx^1\equiv\tilde{\phi}$ while generally it is a function
of $\tx^M$.

Following \cite{Batlle:2016iel,Gomis:2005pg} we define
 non-relativistic limit in the form
\begin{equation}\label{nonreldef}
\tx^M=X^M \ ,\quad  \tx^\mu=\omega X^\mu \ , \quad \mu=0,1 \ , \quad
\ttau_F=\tau_F \  ,
\end{equation}
where $\omega$ is dimensionless parameter that goes to $\infty$
in the non-relativistic limit. Using (\ref{nonreldef}) we find that
$g_{\alpha\beta}$ is equal to
\begin{eqnarray}
g_{\alpha\beta}&=&G_{MN}\partial_\alpha X^M \partial_\beta X^N+
\omega^2\partial_\alpha X^\mu\partial_\beta X^\nu G_{\mu\nu}\equiv \nonumber \\
& \equiv &\omega^2\ba_{\alpha\beta}+
G_{MN}\partial_\alpha X^M\partial_\beta X^N
\nonumber \\
\end{eqnarray}
and hence the action (\ref{NGaction}) takes the form in the limit $\omega\rightarrow \infty$
\begin{eqnarray}\label{Sexpand}
S=-\tau_F\omega^2
\int d\tau d\sigma \sqrt{-\det \ba}-
\frac{\tau_F}{2}\int d\tau d\sigma \sqrt{-\det \ba}\ba^{\alpha\beta}
\partial_\beta X^M\partial_\alpha X^NG_{MN} \
, \nonumber \\
\end{eqnarray}
where the first term diverges in the limit $\omega\rightarrow\infty$. Let us analyze the first term in more details.  Since $\ba_{\alpha\beta}$ is $2\times 2$ matrix and since
$\partial_\alpha X^\mu$ is $2\times 2$ matrix as well we can write
 \begin{equation}
\det \ba=
\det G_{\mu\nu}(\det\partial_\alpha X^\mu)^2 \ ,
\end{equation}
where
\begin{equation}
\det \partial_\alpha X^\mu=
\partial_\tau t\partial_\sigma\phi-\partial_\sigma t\partial_\tau \phi \ .
\end{equation}
Then we can write the divergent contribution in (\ref{Sexpand}) as
\begin{eqnarray}\label{Sdiv}
S_{div}&=&-\tau_F\omega^2\int d\tau d\sigma
\sqrt{-\det G_{\mu\nu}}(\partial_\tau t\partial_\sigma\phi-\partial_\sigma \partial_\tau
\phi)= \nonumber \\
&=&
-\tau_F\omega^2\int d\tau d\sigma
\sqrt{-\det G_{\mu\nu}}
\epsilon^{\alpha\beta}\partial_\alpha t\partial_\beta\phi \ , \nonumber \\
\end{eqnarray}
where $\epsilon^{\tau\sigma}=-\epsilon^{\sigma\tau}=1$. On the other hand
it is well known that fundamental string couples to $NSNS $ two form field
where the coupling term has generally the form
\begin{equation}
S_{NSNS}=\frac{\ttau_F}{2}\int d\tau d\sigma (B_{MN}\epsilon^{\alpha\beta}\partial_\alpha \tx^M
\partial_\beta \tx^N+B_{\mu\nu}\epsilon^{\alpha\beta}\partial_\alpha \tx^\mu
\partial_\beta \tx^\nu) \ .
\end{equation}
Let us now presume that $B_{MN}$ components are zero so that in the non-relativistic limit we find
\begin{equation}\label{NSdiv}
S_{NSNS}=\tau_F\omega^2
\int d\tau d\sigma B_{t\phi}\epsilon^{\alpha\beta}\partial_\alpha t
\partial_\beta \phi \ .
\end{equation}
Now when  $B_{t\phi}=\sqrt{-\det G_{\mu\nu}}$
we find that two divergent contributions (\ref{Sdiv}) and
(\ref{NSdiv}) cancel each other and we are left with the  action for non-relativistic string in the form
\begin{equation}\label{SNRactionfinal}
S_{NR}=-\frac{\tau_F}{2}\int d\tau d\sigma \sqrt{-\det \ba}
\ba^{\alpha\beta} \partial_\beta X^ M \partial_\alpha X^N G_{MN}
\ .
\end{equation}
This action has formally the same form as  Polyakov's  action with the  crucial
difference that the metric $\ba_{\alpha\beta}$ depends on the coordinates
$X^\mu$ and the target space metric $G_{\mu\nu}$ that generally depends on $X^M$.
To see this more explicitly we determine from (\ref{SNRactionfinal})  the equations of motion for $X^M$ and $X^\mu$. In the first case we find
\begin{eqnarray}\label{eqXM}
& &\partial_\alpha[G_{MN}\partial_\beta X^N \ba^{\beta\alpha}
\sqrt{-\det\ba}]-\frac{1}{2}\sqrt{-\det\ba}\ba^{\alpha\beta}\partial_\beta X^K
\partial_\alpha X^L \partial_MG_{KL}-\nonumber \\
&-&T^{\alpha\beta}\sqrt{-\det\ba}\partial_\alpha X^\mu
\partial_M G_{\mu\nu}\partial_\beta X^\nu=0 \ ,
\nonumber \\
\end{eqnarray}
where we  defined $T^{\alpha\beta}(\sigma^\alpha)$ as
\begin{equation}
T^{\alpha\beta}=-\frac{1}{\sqrt{-\det\ba(\sigma^\alpha)}}\frac{\delta S_{NR}}{\delta \ba_{\alpha\beta}(\sigma^\alpha)} \ .
\end{equation}
Using this notation we find that the equations of motion for
 $X^\mu$ have the form
\begin{eqnarray}\label{eqXmu1}
\partial_\alpha[G_{\mu\nu}\partial_\beta X^\nu \sqrt{-\det\ba}T^{\alpha\beta}]=0 \ .
\nonumber \\
\end{eqnarray}
Let us solve the equation (\ref{eqXmu1}) with the ansatz
\begin{equation}\label{Tzero}
T^{\alpha\beta}=0
\end{equation}
even if it is possible that this is too restrictive condition and more general
solutions can be found. Despite of this fact let us analyze consequence
of the equation (\ref{Tzero}).  Explicitly, by definition we have
\begin{equation}
T^{\alpha\beta}=-\frac{1}{2}\ba^{\alpha\beta}\ba^{\gamma\delta}
G_{MN}\partial_\gamma X^M\partial_\delta X^N+\ba^{\alpha\gamma}\ba^{\beta\delta}
G_{MN}\partial_\gamma X^M\partial_\delta X^N \ .
\end{equation}
Now the solutions of the equations $T^{\alpha\beta}=0$ is
\begin{equation}
\ba_{\alpha\beta}=G_{MN}\partial_\alpha X^M\partial_\beta X^N
\end{equation}
as can be easily checked. Inserting this solution into  (\ref{eqXM})
we find that they correspond to the equations of motion derived from
the Nambu-Gotto form of the non-relativistic action $S_{NG}=-
\int d\tau d\sigma \sqrt{-\det G_{MN}\partial_\alpha X^M
\partial_\beta X^N}$.
\section{Hamiltonian Analysis of Non-Relativistic String}\label{third}
Now we perform Hamiltonian analysis of the action
(\ref{SNRactionfinal}), following analysis performed recently in
\cite{Kluson:2017vwp}. We start with $1+1$ decomposition of the matrix
$\ba_{\alpha\beta}$ as it is well known from $3+1$ formalism of General Relativity
\cite{Gourgoulhon:2007ue,Arnowitt:1962hi}
\begin{eqnarray}\label{ba11}
\ba_{\alpha\beta}=\left(\begin{array}{cc}
-n^2+n_\sigma h^{\sigma\sigma}n_\sigma & n_\sigma \\
n_\sigma & h_{\sigma\sigma} \\ \end{array}\right) \ , \quad
\ba^{\alpha\beta}=
\left(\begin{array}{cc}
-\frac{1}{n^2} & \frac{n^\sigma}{n^2} \\
\frac{n^\sigma}{n^2} & h^{\sigma\sigma}-\frac{n^\sigma n^\sigma}{n^2} \\ \end{array}\right) \ ,
\end{eqnarray}
where $n^\sigma=h^{\sigma\sigma}n_\sigma \ , h^{\sigma\sigma}=\frac{1}{h_{\sigma\sigma}}$. Comparing (\ref{ba11}) with
$\ba_{\alpha\beta}=G_{\mu\nu}\partial_\alpha X^\mu\partial_\beta X^\nu$ we find
\begin{eqnarray}
& &h_{\sigma\sigma}=\partial_\sigma X^\mu\partial_\sigma X^\nu G_{\mu\nu}  \ , \quad
n_\sigma=\partial_\tau X^\mu G_{\mu\nu}\partial_\sigma X^\nu \ , \nonumber \\
& & n^2=-\partial_\tau X^\mu (G_{\mu\nu}-G_{\mu\rho}\partial_\sigma X^\rho
h^{\sigma\sigma}\partial_\sigma X^\omega G_{\omega\nu})
\partial_\tau X^\nu\equiv -\partial_\tau
X^\mu V_{\mu\nu}\partial_\tau X^\nu \  \nonumber \\
\end{eqnarray}
and hence the action (\ref{SNRactionfinal}) has the form
\begin{equation}\label{SNRham}
S_{NR}=\frac{\tau_F}{2}\int d\tau d\sigma \sqrt{h_{\sigma\sigma}}n
(\nabla_n X^M\nabla_n X^NG_{MN}-h^{\sigma\sigma}\partial_\sigma X^M\partial_\sigma X^NG_{MN})
\end{equation}
which is suitable for the Hamiltonian analysis. From (\ref{SNRham}) we obtain
 following conjugate momenta
\begin{eqnarray}
p_M&=&\tau_F \sqrt{h_{\sigma\sigma}} G_{MN}\nabla_n X^N \ , \nonumber \\
p_\mu&=&
\frac{\tau_F}{2n}\sqrt{h_{\sigma\sigma}}V_{\mu\nu}\partial_\tau X^\nu[G_{MN}\nabla_n X^M\nabla_n X^N+h^{\sigma\sigma}\partial_\sigma X^M\partial_\sigma X^NG_{MN}]-\nonumber \\
&-&\tau_F\sqrt{h_{\sigma\sigma}}G_{\mu\nu}\partial_
\sigma X^\nu h^{\sigma\sigma}\partial_\sigma X^MG_{MN}
\nabla_n X^N \ . \nonumber \\
\end{eqnarray}
From the last relation we see that it is natural to define object $\omega_\mu$ in the form
\begin{eqnarray}\label{omegadef}
\omega_\mu\equiv p_\mu+ G_{\mu\nu}\partial_\sigma X^\nu h^{\sigma\sigma}\partial_\sigma X^M
p_M=\frac{V_{\mu\nu}\partial_\tau X^\nu}{n}
\left[\frac{p_M G^{MN} p_N}{2\tau_F\sqrt{h_{\sigma\sigma}}}+\frac{\tau_F}{2}\sqrt{h_{\sigma\sigma}}
h^{\sigma\sigma}\partial_\sigma X^M\partial_\sigma X^NG_{MN}\right] \ .
\nonumber \\
\end{eqnarray}
To proceed further we use the fact that
\begin{equation}
V_{\mu\nu}\partial_\sigma X^\nu=0 \ , \quad  V_{\mu\sigma}G^{\sigma\rho}V_{\rho\nu}=
V_{\mu\nu}
 \  .
\end{equation}
which together with (\ref{omegadef}) imply  two primary constraints
\begin{eqnarray}
\mH_\sigma&=&\omega_\mu\partial_\sigma X^\mu=
p_\mu \partial_\sigma X^\mu+p_M\partial_\sigma X^M \approx 0 \ , \nonumber \\
\tmH_\tau&=&\omega_\mu G^{\mu\nu} \omega_\nu+
\frac{1}{\sqrt{h_{\sigma\sigma}}}\left[\frac{1}{2\tau_F}p_M G^{MN}p_N+\frac{\tau_F}{2}\partial_\sigma X^M\partial_\sigma X^NG_{MN}\right]^2\equiv \nonumber \\
&\equiv &
\omega_\mu \omega^\mu+\Sigma^2\approx 0 \ . \nonumber \\
\end{eqnarray}
Since by definition
 $-\omega_\mu G^{\mu\nu}\omega_\nu$ is  positive we can rewrite
$\tmH_\tau$ into the form
\begin{equation}
\tmH_\tau=\Sigma^2-(\sqrt{-\omega_\mu G^{\mu\nu}
\omega_\nu})^2=(\Sigma-\sqrt{-\omega_\mu G^{\mu\nu}
\omega_\nu})(\Sigma+\sqrt{-\omega_\mu G^{\mu\nu}
\omega_\nu})\approx 0 \ .
\end{equation}
Due to the fact that
 $\Sigma$   and $\sqrt{-\omega_\mu G^{\mu\nu}
\omega_\nu}$ are positive definite it is natural to replace $\tmH_\tau$ with
equivalent  primary constraint in the form
\begin{equation}\label{defmHtau}
\mH_\tau=\Sigma-\sqrt{-\omega_\mu G^{\mu\nu}
    \omega_\nu} \approx 0 \ .
\end{equation}
Note that the opposite case when we  would demand that $\Sigma+\sqrt{-\omega_\mu G^{\mu\nu}
    \omega_\nu}$ is a constraint, would be equivalent to the situation when we should impose
 $\Sigma \approx 0$ and $\sqrt{-\omega_\mu G^{\mu\nu}\omega_\nu}\approx 0$ separately
and hence the theory would be over constrained.

Having determined primary constraints of the theory we find that the extended  Hamiltonian with primary constraints included has
the form (Note that the bare Hamiltonian defined as
$H_B=\int d\sigma (p_\mu \partial_\tau X^\mu+p_M\partial_\tau X^M-\mL)$ is zero.)
\begin{equation}\label{extHam}
H_E=\int d\sigma (\lambda^\tau \mH_\tau+\lambda^\sigma \mH_\sigma) \ .
\end{equation}
As the next step we determine Poisson bracket between smeared form of the constraints
$\mH_\tau,\mH_\sigma$ defined as
\begin{equation}
\bT_\tau(N)=\int d\sigma N \mH_\tau \ , \quad
\bT_\sigma(N^\sigma)=\int d\sigma N^\sigma \mH_\sigma \ .
\end{equation}
First of all it  is easy to see that
\begin{eqnarray}
\pb{\bT_\sigma(N^\sigma),\bT_\sigma(M^\sigma)}=
\bT_\sigma(N^\sigma \partial_\sigma M^\sigma-M^\sigma \partial_\sigma N^\sigma) \ .
\nonumber \\
\end{eqnarray}
To proceed further  we need Poisson bracket between
$\bT_\sigma(N^\sigma)$ and rest of canonical variables. It is useful
to write $\omega_\mu$ as
\begin{equation}
\omega_\mu=p_\mu+G_{\mu\nu}\partial_\sigma X^\nu h^{\sigma\sigma}
\partial_\sigma X^M p_M=(\delta_\mu^\rho-G_{\mu\nu}
\partial_\sigma X^\nu h^{\sigma\sigma}\partial_\sigma X^\rho)p_\rho+
G_{\mu\nu}\partial_\sigma X^\nu h^{\sigma\sigma}\mH_\sigma\approx
V_\mu^{ \ \rho} p_\rho
\end{equation}
and hence we find
\begin{eqnarray}
& &\pb{\bT_\sigma(N^\sigma),h_{\sigma\sigma}}=-N^\sigma
\partial_\sigma h_{\sigma\sigma}-2\partial_\sigma N^\sigma h_{\sigma\sigma} \ , \nonumber \\
& &\pb{\bT_\sigma(N^\sigma),h^{\sigma\sigma}}=-N^\sigma \partial_\sigma h^{\sigma\sigma}+
2\partial_\sigma N^\sigma h^{\sigma\sigma}
 \ , \nonumber \\
& &\pb{\bT_\sigma(N^\sigma),\omega_\mu}=-\partial_\sigma (N^\sigma \omega_\mu) \
\nonumber \\
\end{eqnarray}
so that
\begin{eqnarray}
\pb{\bT_\sigma(N^\sigma),\bT_\tau(M)}=\bT_\tau(N^\sigma\partial_\sigma M) \ .
 \nonumber \\
\end{eqnarray}
Finally we calculate the Poisson bracket between $\bT_\tau(N),\bT_\tau(M)$.
Following \cite{Kluson:2017vwp} and after some calculations we find
\begin{eqnarray}
\pb{\bT_\tau(N),\bT_\tau(M)}=
\bT_\sigma((N\partial_\sigma M-M\partial_\sigma N)h^{\sigma\sigma}) \ .
\nonumber \\
\end{eqnarray}
These results show that the constraints $\mH_\tau
\approx 0 \ , \mH_\sigma\approx 0$ are the first class constraints
that reflect two dimensional diffeomorphism invariance of the non-relativistic
string world-sheet.
Before we proceed to the analysis of gauge fixed theory we
determine equations of motion using an extended Hamiltonian $H_E$. Note
that all metric components generally depend on $X^M$ but do not dependent on $X^\mu$. Then we obtain
\begin{eqnarray}
& &\partial_\tau X^\mu=\pb{X^\mu,H_E}=\lambda^\tau \frac{V^{\mu\nu}p_\nu}
{\sqrt{-\omega_\mu \omega^\mu}}+\lambda^\sigma \partial_\sigma x^\mu \ , \nonumber \\
& &\partial_\tau p_\mu=
\nonumber \\
&=&\partial_\sigma\left[\frac{\lambda^\tau}{h_{\sigma\sigma}\sqrt{-\omega_\mu
    \omega^\mu}}(-p_\mu \partial_\sigma X^\rho p_\rho+G_{\mu\sigma}
\partial_\sigma X^\sigma p_\rho G^{\rho\sigma}p_\sigma)
-\lambda^\tau \frac{G_{\mu\sigma}\partial_\sigma X^\sigma}{h_{\sigma\sigma}}\mH_\tau+\lambda^\sigma p_\mu
\right]\approx
\nonumber \\
&\approx &
\partial_\sigma\left[\frac{\lambda^\tau}{h_{\sigma\sigma}\sqrt{-\omega_\mu
        \omega^\mu}}(-p_\mu \partial_\sigma X^\rho p_\rho+G_{\mu\sigma}
\partial_\sigma X^\sigma p_\rho G^{\rho\sigma}p_\sigma)
+\lambda^\sigma p_\mu
\right] \ , \nonumber \\
\end{eqnarray}
where
\begin{equation}
V^{\rho\sigma}=G^{\rho\sigma}-\partial_\sigma X^\rho h^{\sigma\sigma}
\partial_\sigma X^\sigma \ .
\end{equation}
The equations of motion  for $X^M$ have the form
\begin{eqnarray}\label{eqXMham}
\partial_\tau X^M=\pb{X^M,H_E}=\frac{\lambda^\tau }{\tau_F \sqrt{h_{\sigma\sigma}}}
G^{MN}p_N+\lambda^\sigma \partial_\sigma X^M \ . \nonumber \\
\end{eqnarray}
In case of the equations of motion for $p_M$ we would get much more complicated
result due to the explicit dependence of $V^{\mu\nu},h_{\sigma\sigma}$ on $X^M$. Lucky we will not need them in what follows.
\section{Uniform Gauge for Non-Relativistic String}\label{fourth}
In this section we find gauge fixed Hamiltonian for non-relativistic string
in uniform gauge, following mainly
\cite{Arutyunov:2004yx,Arutyunov:2013ega,Arutynov:2014ota} when
we presume that
the background metric has the form
\begin{eqnarray}
ds^2&=&G_{tt}dt^2+G_{\phi\phi}d\phi^2+G_{MN}dx^M dx^N \ . \nonumber \\
\end{eqnarray}
Following
\cite{Arutyunov:2013ega,Arutynov:2014ota} we introduce light cone
coordinates
\begin{equation}
x^-=\phi-t \ , \quad x^+=(1-a)t+a\phi \
\end{equation}
with inverse relations
\begin{equation}
t=x^+-ax^- \ , \quad \phi=x^++(1-a)x^- \ ,
\end{equation}
where $a$ is a free parameter $a\in (0,1)$. Using this definition it
is easy to find following components of  metric
\begin{equation}
G_{++}=G_{tt}+G_{\phi\phi} \ , \quad
G_{--}=G_{tt}a^2+(1-a)^2G_{\phi\phi} \ , \quad  G_{+-}=
-aG_{tt}+(1-a)G_{\phi\phi} \
\end{equation}
with inverse
\begin{eqnarray}
G^{++}&=&
\frac{G_{tt}a^2+(1-a)^2G_{\phi\phi}}{G_{tt}G_{\phi\phi}} \ , \quad
G^{--}=
\frac{G_{tt}+G_{\phi
        \phi}}{G_{tt}G_{\phi\phi}} \ ,
\nonumber \\
G^{+-}&=&
\frac{aG_{tt}-(1-a)G_{\varphi\varphi}}{G_{tt}G_{\varphi\varphi}} \
.
\nonumber \\
\end{eqnarray}
Now we are ready to impose the uniform  light cone gauge fixing
\cite{Arutyunov:2004yx,Arutyunov:2013ega,Arutynov:2014ota}
when we introduce
two gauge fixing constraints
\begin{equation}
\mG_+\equiv x^+-\tau \approx  0 \ , \quad \mG_-\equiv  p_--J\approx
0 \ .
\end{equation}
It is also important to stress that we have to ensure the
preservation of these constrains during the time evolution of the
system. Explicitly, the extended Hamiltonian with primary  constraints
and gauge fixing constraints included has the form
\begin{equation}
H_T=\int d\sigma (\lambda^\tau \mH_\tau+\lambda^\sigma \mH_\sigma
+u_+\mG_++u_-\mG_-) \ .
\end{equation}
Now the requirement of the preservation of the constraint $\mG_+
\approx 0$ implies  following equation for
$\lambda^\tau$
\begin{eqnarray}
\frac{d}{d\tau} \mG_+&=&\frac{\partial \mG_+}{\partial t}+\pb{\mG_+,H_T}=
-1+\lambda^\tau \frac{V^{+\nu}p_\nu}{\sqrt{-\omega_\mu G^{\mu\nu}
\omega_\nu}}
+\lambda^\sigma \partial_\sigma x^+=\nonumber \\
&=&-1+\lambda^\tau \frac{1}{\sqrt{-(\omega_\mu G^{\mu\nu}\omega_\nu)}}(V^{++}p_++V^{+-}p_-)
\
\nonumber \\
\end{eqnarray}
that can be solved for $\lambda^\tau$ as
\begin{equation}\label{taudet}
\lambda^\tau=\frac{\sqrt{-\omega_\mu G^{\mu\nu}\omega_\nu}}{(V^{++}p_++V^{+-}p_-)}=
\frac{\Sigma}{V^{++}p_++V^{+-}p_-} \ ,
\end{equation}
where in the second step we used the constraint $\mH_\tau=0$.
As the next step we determine the time evolution of the constraint
$\mG_-$
\begin{eqnarray}\label{dmGminus}
\frac{d}{d\tau} \mG_-=\pb{\mG_-,H_T}=
\partial_\sigma \left[\frac{p_M\partial_\sigma X^M}{J(V^{++}p_++V^{+-}p_-)h_{\sigma\sigma}}\Sigma+\lambda^\sigma J\right]=0 \nonumber \\
\end{eqnarray}
using following Poisson bracket
\begin{equation}
\pb{p_-(\sigma),\Sigma(\sigma')}=\frac{1}{h_{\sigma\sigma}(\sigma')}\Sigma(\sigma') \partial_{\sigma'} \delta(\sigma-\sigma')G_{--}(\sigma')\partial_{\sigma'}  X^-(\sigma')  \ .
\end{equation}
The equation (\ref{dmGminus}) can be solved for $\lambda^\sigma$ as
\begin{equation}\label{sigmadet}
\lambda^\sigma(\tau,\sigma)=
-\frac{p_M\partial_\sigma X^MG_{--}}{J^2h_{\sigma\sigma}(V^{++}p_++V^{+-}p_-)}
\end{equation}
up to an
arbitrary function
$f(\tau)$.  The fact, that the Lagrange multiplier $\lambda^\sigma$ is fixed
up to the function $f(\tau)$ suggests an existence of the global constraint
defined as
\begin{equation}
\bT_\sigma=\int d\sigma \mH_\sigma \ .
\end{equation}
Then we have to separate this constraint from the local one
$\mH_\sigma$. To do this we define following local constraint
\begin{equation}
\tmH_\sigma=\mH_\sigma-\frac{\sqrt{h_{\sigma\sigma}}}{\int d\sigma'
    \sqrt{h_{\sigma\sigma}}} \bT_\sigma \  ,
\end{equation}
where  $\tmH_\sigma$ is restricted since
\begin{equation}
\int d\sigma \tmH_\sigma=0 \ .
\end{equation}
Then it is clear that we can write the contribution of the
diffeomorphism constraint to the total Hamiltonian in the form
\begin{equation}
\int d\sigma (\tlambda^\sigma \tmH_\sigma)+f(\tau)\bT_\sigma  \ ¨,
\end{equation}
where again by definition $\int d\sigma \tlambda^\sigma=0$
so that the requirement of the preservation of the constraint
$\mG_-$ implies
 $\tlambda^\sigma=0$. It is also clear that $\mG_-\approx 0$
does not fix the constraint $\bT_\sigma$ since
\begin{equation}
\pb{\mG_-(\sigma),\bT_\sigma}=\partial_\sigma p_-=0 \
\end{equation}
using the fact that
 $\bT_\sigma$ is the generator of the global
transformations
\begin{equation}
\pb{p_M,\bT_\sigma}=\partial_\sigma p_M \ , \quad
\pb{X^M,\bT_\sigma}=\partial_\sigma X^M \ .
\end{equation}
As the final step we analyze the requirement of the preservation of the constraints
$\mH_\tau,\tmH_\sigma$. We begin with $\tmH_\sigma$ and we obtain
\begin{eqnarray}
\partial_\tau \mH_\sigma=\pb{\mH_\sigma,H_E}=
- u^+\partial_\sigma X^+-\partial_\sigma (p_-u^-)\approx
-J\partial_\sigma (u^-)=0 \nonumber \\
\end{eqnarray}
that implies that $u^-=0$ since we presume that this Lagrange multiplier does not include zero mode since $\mG_-$  fixes the constraint $\tmH_\sigma$. Using this result we easily determine the equation of motion for $\mH_\tau\approx 0$
\begin{eqnarray}
\partial_\tau \mH_\tau&=&\pb{\mH_\tau,H_E}\approx  \int d\sigma
u_+(\sigma)\pb{\mH_\tau,\mG_+(\sigma)}=0 \   \nonumber \\
\end{eqnarray}
that implies $u_+=0$.

Let us now determine Hamiltonian on the reduced phase space.
Due to the gauge fixing we find that
$\mH_\tau,\tmH_\sigma$ vanishes strongly so that the action has the
form
\begin{eqnarray}
S&=&\int d\tau d\sigma (p_\mu\partial_\tau X^\mu+p_M\partial_\tau X^M-\lambda_\tau \mH_\tau-
\tlambda_\sigma \tmH_\sigma)-\int d\tau f(\tau)\bT_\sigma= \nonumber
\\
&=&\int d\tau d\sigma (p_M \partial_\tau X^M+p_+)-\int d\tau f(\tau)
\bT_\sigma \nonumber \\
\end{eqnarray}
and we see that it is natural to identify $-p_+$ as the Hamiltonian
density for the reduced phase space. We express the dependence $p_+$
on $p_M$ when we solve the constraints $\mH_\tau=0,\tmH_\sigma=0$
\begin{eqnarray}
p_+=\frac{-V^{+-}\pm \sqrt{V^{+-}V^{+-}J^2-V^{++}(J^2 V^{--}+\Sigma^2)}}
{V^{++}}  \nonumber \\
\end{eqnarray}
so that the reduced Hamiltonian has the form
\begin{eqnarray}\label{Hred}
H_{red}=\int d\sigma \left(
\frac{V^{+-}}{V^{++}}J-\frac{1}{V^{++}} \sqrt{V^{+-}V^{+-}J^2-V^{++}(J^2 V^{--}+\Sigma^2)} \right)+f(\tau)\bT_\sigma \ ,
\nonumber \\
\end{eqnarray}
where we have chosen $+$ sign in front of the square root in order to have positive definite Hamiltonian
The Hamiltonian (\ref{Hred}) seems to be rather complicated expression. First of all we see that it is a function of  $\Sigma^2$
rather than $\Sigma $ as in relativistic case
\cite{Arutyunov:2004yx}. Further, $\Sigma$ depends on $h_{\sigma\sigma}$ that on the constraint surface is equal to
\begin{eqnarray}
h_{\sigma\sigma}
=\partial_\sigma X^-\partial_\sigma X^-G_{--}=
\frac{1}{J^2}(p_M\partial_\sigma X^M)^2G_{--} \ .
\end{eqnarray}
We also find that on the constraint surface the projectors
$V^{++},V^{+-}$ and $V^{--}$ are equal to
\begin{eqnarray}
V^{++}&=&G^{++} \ ,  \quad
V^{+-}=G^{+-} \ , \quad
V^{--}=
\frac{G^{--}G_{--}-1}{G_{--}} \ .
\nonumber \\
\end{eqnarray}
Using the Hamiltonian on the reduced phase space we derive equations of motion for $X^M$ as
\begin{eqnarray}\label{eqxMred}
\partial_\tau X^M=\pb{X^M,H_{red}}=\frac{1}{\sqrt{\bK}\tau_F\sqrt{h_{\sigma\sigma}}}G^{MN}p_N\Sigma-
\frac{1}{J^2h_{\sigma\sigma}\sqrt{\bK}}
\partial_\sigma X^M G_{--}(p_N\partial_\sigma X^N)
\Sigma \ , \nonumber \\
\end{eqnarray}
where
\begin{equation}
\bK=(V^{+-}J)^2-V^{++} (V^{--}J^2+\Sigma^2) \ .
\end{equation}
As a check, we can compare this equation of motion with the equations of motion
given in (\ref{eqXMham}) using the values of the Lagrange multipliers
$\lambda^\tau,\tlambda^\sigma$ determined  by gauge fixing procedure. Explicitly, inserting (\ref{taudet}) and (\ref{sigmadet}) into (\ref{eqXMham}) we obtain
\begin{eqnarray}\label{eqlambda}
\partial_\tau X^M
&=&\frac{\Sigma}{\tau_F(G^{++}p_++G^{+-}p_-)\sqrt{h_{\sigma\sigma}}}
G^{MN}p_N
-\frac{\Sigma p_N\partial_\sigma X^NG_{--}}{J^2 (G^{++}p_++G^{+-}p_-)h_{\sigma\sigma}}\partial_
\sigma X^M  \nonumber \\
&=&\frac{1}{\sqrt{\bK}\tau_F\sqrt{h_{\sigma\sigma}}}
G^{MN}p_N\Sigma
-\frac{1}{J^2h_{\sigma\sigma} \sqrt{\bK}}\partial_
\sigma X^M G_{--}
\Sigma (p_N\partial_\sigma X^N)
\ , \nonumber \\
\end{eqnarray}
using
\begin{equation}
G^{++}p_++G^{+-}p_-=\sqrt{\bK} \ .
\end{equation}
We see that (\ref{eqlambda}) coincide with (\ref{eqxMred}) which is a nice
check of our procedure.
We can use these equations of motion to invert Hamiltonian on the reduced phase space to corresponding Lagrangian density for the physical degrees
of freedom and we can analyze its properties. In particular, it would be interesting
to study equations of motion that follow from this Lagrangian and
try to find solutions that have similar properties as giant magnon
\cite{Hofman:2006xt}.
We hope to return to this problem in  future.
\section{Integrability of Non-Relativistic String}\label{fifth}
Let us now discuss the question of the integrability of the non-relativistic string. As the first case we consider situation when the target space-time
metric has the form
\begin{equation}
ds^2=\eta_{\mu\nu}dX^\mu dX^\nu+G_{MN}dX^M dX^N \ ,
\end{equation}
where
\begin{equation}
G_{MN}=E_M^{ \ A}E_N^{ \  B}K_{AB} \ , \quad  K_{AB}=\tr (T_A T_B) \ ,
\end{equation}
where
\begin{equation}
g^{-1}\partial_\alpha g=g^{-1}\partial_M g \partial_\alpha X^M=
E_M^{ \ A}T_A \partial_\alpha X^M \ ,
\end{equation}
and where by definition $E_M^{ \ A}$ obey the equation
\begin{equation}\label{flatproperty}
\partial_M E_N^{ \ A}-\partial_N E_M^{\ A}+E_M^{\  B} E_N^{ \
C}f_{BC}^{ \ \quad A}=0 \ ,
\end{equation}
where $g$ is the group element from some group $F$ that defines group manifold with the metric $G_{MN}$ given above. $T_A$ are basis elements of the Lie algebra of this group with the corresponding metric $K_{AB}$.  Finally the constants $f_{AB}^{ \ \quad C}$ are defined as $[T_A,T_B]=f_{AB}^{ \ \quad C}T_C$.

Since the metric $G_{\mu\nu}=\eta_{\mu\nu}$ is flat we find that the background
NSNS two form  that is needed for the
cancelation of the divergent contribution is  equal to
\begin{equation}
B_{t\phi}=1 \
\end{equation}
that clearly solves the background equations of motion and hence non-relativistic
string can be consistently defined. Then we  propose
following Lax connection in the form
\begin{eqnarray}\label{Laxconflat}
L_\sigma(\Lambda)&=&\frac{1}{1-\Lambda^2}[E_M^{ \ A}\partial_\sigma
X^M+\Lambda \ba^{\tau\alpha}E_M^{ \ A}\partial_\alpha X^M\sqrt{-\det
\ba}] \ , \nonumber \\
L_\tau(\Lambda)&=&\frac{1}{1-\Lambda^2}[E_M^{ \ A}\partial_\tau X^M-
\Lambda \ba^{\sigma\alpha}E_M^{ \ A}\partial_\alpha X^M \sqrt{-\det
    \ba}] \  \nonumber \\
\end{eqnarray}
and we are going to argue that it is flat. As the first step we show that the
equation of motion for $X^M$ have the form
\begin{eqnarray}
K_{AB}E_M^{ \ A}\partial_\alpha [E_N^{ \ B}\partial_\beta X^N
\ba^{\beta\alpha}\sqrt{-\det \ba}]=0 \ , \nonumber \\
\end{eqnarray}
using (\ref{flatproperty}) and the fact that $\ba_{\alpha\beta}$
does not depend on $X^M$. Now due to the invertibility of $E_M^{ \ A}$
we see that the equations of motion for $X^M$ can be written into
an equivalent form
\begin{equation}
\partial_\alpha[E_M^{ \ A}\partial_\beta X^M \ba^{\beta\alpha}
\sqrt{-\det \ba}]=0 \
\end{equation}
and hence we easily find that the Lax connection (\ref{Laxconflat})
is flat
\begin{equation}
\partial_\tau L_\sigma^A-\partial_\sigma L_\tau^A+L_\tau^{ \ B}
L_\sigma^{ \ C}f_{BC}^{ \ \quad A}=0 \ .
\end{equation}
We see that the theory is trivially integrable since the non-relativistic
limit was performed in the flat space. It is more interesting to consider
the case when the whole space-time corresponds to group manifold as for example
$AdS_5\times S^5$ space-time.
\subsection{Non-Relativistic Limit of String on $AdS_5\times S^5$}\label{sixth}
More interesting situation occurs when we analyze string on $AdS_5\times S^5$. To begin with we introduce notations and conventions, following very nice paper \cite{Arutyunov:2004yx}.
The space time $AdS_5\times S^5$ is a coset
\begin{equation}
AdS_5\times S^5=SO(2,4)\times SO(6)/SO(5,1)\times SO(5)
\end{equation}
so that the string sigma model corresponds to sigma models on group and coset manifolds. We describe this coset using coset representative where the matrix $g$
has the form
\begin{equation}
g=\left(\begin{array}{cc}
g_a & 0 \\
0 & g_s \\
\end{array}\right) \ .
\end{equation}
Here $g_a$ and $g_s$ are  following $4\times 4$ matrices
\begin{equation}
g_a=\left(\begin{array}{cccc}
0 & \mZ_3 & -\mZ_2 &\mZ_1^* \\
-\mZ_3& 0 & \mZ_1 & \mZ_2^* \\
\mZ_2 & -\mZ_1 & 0 &-\mZ_3^* \\
-\mZ_1^* & -\mZ_2^* & \mZ_3^*
&0 \\
\end{array}\right) \ , \quad
g_s=\left(\begin{array}{cccc}
0 & \mY_1 & -\mY_2 &\mY_3^* \\
-\mY_1& 0 & \mY_3 & \mY_2^* \\
\mY_2 & -\mY_3 & 0 & \mY_1^* \\
-\mY_3^* & -\mY_2^* & -\mY_1^*
&0 \\
\end{array}\right) \ ,
\end{equation}
where $\mZ_k, k=1,2,3$ are the complex embedding coordinates for
$AdS_5$ and $\mY_k \ , k=1,2,3$ are the complex embedding
coordinates for sphere.  The matrix $g_a$ is an element of the group
$SU(2,2)$ since it can be shown that
\begin{equation}
g_a^\dag E g_a=E \ , \quad E=\mathrm{diag}(-1,-1,1,1)
\end{equation}
provided the following condition is satisfied
\begin{equation}
\mZ_1^*\mZ_1+\mZ_2^*\mZ_2 -\mZ_3^*\mZ_3=-1 \ .
\end{equation}
In fact $g_a$ describes embedding of an element of the coset space
$SO(4,2)/SO(5,1)$ into group $SU(2,2)$ that is locally isomorphic to
$SO(4,2)$. We use this isometry to work with $4\times 4$ matrices
rather with $6\times 6$ ones. Note that due to the explicit choice
of the coset representative above there is not any gauge symmetry
left. Quite analogously $g_s$ is unitary
\begin{equation}
g_s g_s^\dag=\mathrm{I} \ , \quad \mathrm{I}=\mathrm{diag}(1,1,1,1)
\end{equation}
on condition that $\mY_1^*\mY_1+ \mY_2^*\mY_2+\mY^*_3\mY_3=1$. The
matrix $g_s$ describes an embedding of an element of the coset
$SO(6)/SO(5)$ into $SU(4)$ being isomorphic to $SO(6)$.

The variables $\mZ,\mY$ are related to the variables $\tx^M$ and $\tx^\mu$ in the following way.
 The five sphere $S^5$ is
parameterized by five variables: coordinates $y^i \ , i=1,\dots,4$
and the angle variable $\phi$. In terms of six real embedding
coordinates $Y^A \ , A=1,\dots,6$ obeying the condition $Y_AY^A=1$
the parametrization reads
\begin{eqnarray}
\mY_1&=&Y_1+iY_2=\frac{y_1+iy_2}{ 1+\frac{y^2}{4}} \ , \quad
\mY_2=Y_3+iY_4=\frac{y_3+iy_4}{ 1+\frac{y^2}{4}} \ ,
\nonumber \\
\mY_3&=&Y_5+iY_6=\frac{1-\frac{y^2}{4}}{ 1+\frac{y^2}{4}}\exp
(i\phi)
 \ .
\nonumber \\
\end{eqnarray}
In the same way we describe the $AdS_5$ space when we introduce four
coordinates $z_i$ and $t$. The embedding coordinates $Z_A$ that obey
$Z_AZ_B\eta^{AB}=-1$ with the metric $\eta^{AB}=(-1,1,1,1,1,-1)$ is
now parameterized as
\begin{eqnarray}
\mZ_1&=&Z_1+iZ_2=-\frac{z_1+iz_2}{ 1-\frac{z^2}{4}} \ , \quad
\mZ_2=Z_3+iZ_4=-\frac{z_3+iz_4}{ 1-\frac{z^2}{4}} \ ,
\nonumber \\
\mZ_3&=&Z_0+iZ_5=\frac{1+\frac{z^2}{4}}{ 1-\frac{z^2}{4}}\exp (it)
 \ .
\nonumber \\
\end{eqnarray}
Note that the line element for $AdS_5\times S^5$ takes the form
\begin{eqnarray}
ds^2=-\frac{(1+\frac{z^2}{4})^2}{(1-\frac{z^2}{4})^2}dt^2
+\frac{1}{(1-\frac{z^2}{4})^2}dz_idz_i+
\left(\frac{1-\frac{y^2}{4}}{1+\frac{y^2}{4}}\right)^2d\phi^2+
\frac{1}{(1+\frac{y^2}{4})^2}dy_idy_i \ .
\nonumber \\
\end{eqnarray}
In other words, $\tx^M=(y_i,z_i)$.
Our goal is to find Lax connection for non-relativistic theory where
we define non-relativistic string performing limiting procedure
 along $t$ and $\phi$ coordinates. Following general discussion in the second
 section this fact implies that
there should be non-zero two form
\begin{equation}
    B_{t\phi}=\sqrt{-\det G_{\mu\nu}}=
    \sqrt{-G_{tt}G_{\phi\phi}}=
    \frac{(1+\frac{z^4}{4})(1-\frac{y^2}{4})}
    {(1-\frac{z^2}{4})(1+\frac{y^2}{4})} \ .
\end{equation}
It is an open question whether such a background $B$ field solves
the supergaravity equations of motion. Despite of this fact let us proceed
further and try to analyze flat connection for this string.
%
In order to find non-relativistic string on this background we proceed as in
 \cite{Frolov:2005dj,Arutyunov:2004yx}
and use the fact  that
 $g_s,g_a$ enjoy following
property
\begin{eqnarray}
g_s(y,\tphi)&=& M(\tphi)\hg_s(y)M(\tphi)
\ , \nonumber \\
g_a(z,\tit)&=& N(t)\hg_a(z) N(\tit) \ ,
\nonumber \\
\end{eqnarray}
where
\begin{equation}
M(\phi)= \left(\begin{array}{cccc}
e^{-\frac{i}{2}\phi} & 0 & 0 & 0 \\
0 & e^{\frac{i}{2}\phi} & 0 & 0 \\
0 & 0 & e^{\frac{i}{2}\phi} & 0 \\
0 & 0 & 0 & e^{-\frac{i}{2}\phi} \\
\end{array}\right) \ , \quad
N(t)= \left(\begin{array}{cccc}
e^{\frac{i}{2}t} & 0 & 0 & 0 \\
0 & e^{\frac{i}{2}t} & 0 & 0 \\
0 & 0 & e^{-\frac{i}{2}t} & 0 \\
0 & 0 & 0 & e^{-\frac{i}{2}t} \\
\end{array}\right) \ ,
\end{equation}
and
\begin{equation}
\hg_a=\left(\begin{array}{cccc} 0 & \frac{1+\frac{z^2}{4}}
{1-\frac{z^2}{4}} & -\mZ_2 &\mZ_1^* \\
-\frac{1+\frac{z^2}{4}} {1-\frac{z^2}{4}}
& 0 & \mZ_1 & \mZ_2^* \\
\mZ_2 & -\mZ_1 & 0 &- \frac{1+\frac{z^2}{4}}
{1-\frac{z^2}{4}} \\
-\mZ_1^* & -\mZ_2^* & \frac{1+\frac{z^2}{4}} {1-\frac{z^2}{4}}
&0 \\
\end{array}\right) \ , \quad
\hg_s=\left(\begin{array}{cccc} 0 & \mY_1 & -\mY_2 &
\frac{1-\frac{y^2}{4}}
{1+\frac{y^2}{4}}  \\
-\mY_1& 0 & \frac{1-\frac{y^2}{4}}
{1+\frac{y^2}{4}} & \mY_2^* \\
\mY_2 & - \frac{1-\frac{y^2}{4}} {1+\frac{y^2}{4}}
 & 0 & \mY_1^* \\
-\frac{1-\frac{y^2}{4}} {1+\frac{y^2}{4}} & -\mY_2^* & -\mY_1^*
&0 \\
\end{array}\right) \ .
\end{equation}
Since $N^\dag E N=E$ we immediately find $\hg^{\dag}_aE\hg_a=E$.
In the same way we
find that $\hg_s^{\dag}\hg_s=\mathrm{I}$.
Note that in this case the matrix $g$ can be written as
\begin{equation}\label{Gmhg}
g=\bM\hg \bM \ , \quad \bM=\left(\begin{array}{cc}
N(t) & 0 \\
0 & M(\phi) \\ \end{array} \right)
 \ , \quad
 \hg=\left(\begin{array}{cc}
 \hg_a & 0 \\
 0 & \hg_s \\ \end{array}\right) \ ,
 \end{equation}
 where crucially $\bM$ does not depend on $\tx^M$.
Using this fact we now determine components of the metric $G_{MN}$ as
\begin{eqnarray}
G_{MN}=\tr (g^{-1}\partial_M g g^{-1}\partial_N g)
=\tr (\hg^{-1}\partial_M \hg\hg^{-1}\partial_N \hg) \nonumber \\
\end{eqnarray}
using the fact that $\bM$ does not depend on $\tx^M$. Then we can
write
\begin{eqnarray}
& &\det (G_{\mu\nu}\partial_\alpha\tx^\mu\partial_\beta\tx^\nu
+G_{MN}\partial_\alpha \tx^M
\partial_\beta \tx^N)=\nonumber \\
&=&\det (\omega^2\ba_{\alpha\beta}+\tr  (\hg^{-1}\partial_M \hg\hg^{-1}\partial_N \hg) \partial_\alpha X^M\partial_\beta X^N)=
\nonumber \\
&=&\det (\omega^2\ba_{\alpha\beta}+\tr \hJ_\alpha\hJ_\beta) \nonumber \\
\end{eqnarray}
and hence non-relativistic limit leads to the action
\begin{equation}\label{nonrelAdS}
S_{NF}=-\frac{\tau_F}{2}\int d\tau d\sigma \sqrt{-\det \ba}
\ba^{\alpha\beta}\tr \hJ_\alpha \hJ_\beta  \ , \quad  \hJ_\alpha=
\hg^{-1}\partial_M \hg \partial_\alpha X^M \ .
\end{equation}
Since (\ref{nonrelAdS}) formally looks as Polyakov's form of string action on
group manifold  it
is natural to propose
 Lax connection in the form
\begin{eqnarray}
L_\sigma(\Lambda)=\frac{1}{1-\Lambda^2}[\hJ_\sigma
+\Lambda \ba^{\tau\alpha}\hJ_\alpha \sqrt{-\det
    \ba}] \ , \nonumber \\
L_\tau(\Lambda)=\frac{1}{1-\Lambda^2}[\hJ_\tau-
\Lambda \ba^{\sigma\alpha}\hJ_\alpha \sqrt{-\det
    \ba}] \ . \nonumber \\
\end{eqnarray}
To proceed further we determine equations of motion for $X^M$
\begin{eqnarray}
\hE_M^{ \ A}K_{AB}\partial_\alpha[\hJ^B_\beta
\ba^{\beta\alpha}\sqrt{-\det \ba}]-
T^{\alpha\beta}\sqrt{-\det\ba}\partial_\alpha X^\mu\partial_M G_{\mu\nu}
\partial_\beta X^\nu=0 \nonumber \\
\end{eqnarray}
that can be rewritten into an equivalent form
\begin{eqnarray}
\partial_\alpha[\hJ_\beta^A\ba^{\beta\alpha}\sqrt{-\det\ba}] -
K^{AB}\hE_B^{ \ M}T^{\alpha\beta}\partial_M G_{\mu\nu}
\partial_\alpha X^\mu\partial_\beta X^\nu=0 \ , \nonumber \\
\end{eqnarray}
where we used the fact that we can write
\begin{eqnarray}
\hg^{-1}\partial_M \hg=\hE_M^{ \ A}T_A \ ,  \quad
G_{MN}
=\hE_M^{ \ A}\hE_N^{ \ B}K_{AB} \  , \quad  K_{AB}=\tr (T_A T_B) \ ,
\nonumber \\
\end{eqnarray}
where by definition
\begin{eqnarray}
\partial_M \hE_N^{ \ A}-\partial_N \hE_M^{ \ A}+
\hE_M^{ \ B}\hE_N^{ \ C}f_{BC}^{\ \quad  A}=0  \ .
\end{eqnarray}
Then it is easy to see that
\begin{eqnarray}
\partial_\tau L_\sigma^A-\partial_\sigma L_\tau^A
=-\frac{1}{1-\Lambda^2}\hJ^B_\tau \hJ^C_\sigma f_{BC}^{ \ A}+
\frac{\Lambda}{1-\Lambda^2}K^{AB}\hE_B^{ \ M}T^{\alpha\beta}
\partial_M G_{\mu\nu}\partial_\alpha X^\mu \partial_\beta X^\nu \ .  \nonumber \\
\end{eqnarray}
On the other hand we have
\begin{eqnarray}
L^B_\tau L^C_\sigma f_{BC}^{  \ \quad  A}=
\frac{1}{1-\Lambda^2}\hJ^B_\tau \hJ^C_\sigma f_{BC}^{ \ \quad A} \nonumber \\
\end{eqnarray}
so that together we obtain
\begin{eqnarray}
\partial_\tau L_\sigma^A-\partial_\sigma L_\tau^A+L_\tau^B L_\sigma^C f_{BC}^{ \ \quad A}=
\frac{\Lambda}{1-\Lambda^2}K^{AB}\hE_B^{ \ M}T^{\alpha\beta}
\partial_M G_{\mu\nu}\partial_\alpha X^\mu \partial_\beta X^\nu \ .  \nonumber \\
\end{eqnarray}
We see that this current is flat when we solve the equations of motion for $X^\mu$ by the ansatz $T^{\alpha\beta}=0$. Note that this condition however implies
\begin{equation}
\ba_{\alpha\beta}=\tr \hJ_\alpha \hJ_\beta
\end{equation}
We have shown that the Lax connection $L^A_\alpha$ is flat.  On the other hand
the existence of flat Lax connection is necessary condition of integrability but not
sufficient since the theory  is integrable if there is an infinite number of charges in involution, see for example \cite{Dorey:2006mx}. Since these charges are defined using monodromy matrix and since this matrix is constructed with the help of the spatial components of Lax connection it is sufficient to calculate Poisson brackets between the spatial components of Lax connection for two different spectral parameters. To do this  we have to  express the spatial component of Lax connection for non-relativistic
string using canonical variables
\begin{eqnarray}\label{Laxspatialcan}
L^A_\sigma=\frac{1}{1-\Lambda^2}[\hE_M^{ \ A}\partial_\sigma X^M-
\frac{1}{\tau_F}\Lambda \hE_M^{ \ A} G^{MN}p_N]  \ .
\nonumber \\
\end{eqnarray}
We see very interesting fact that now this component does not depend on matrix  $\ba_{\alpha\beta}$
and hence does not depend on $X^\mu$. This is in sharp contrast with the time component of the Lax connection where in order to find its canonical form
we have to use   equations of motion in order to replace time derivatives
with corresponding Poisson brackets between these variables and Hamiltonian.
Now using (\ref{Laxspatialcan})  we  easily determine corresponding
Poisson brackets between spatial component of the Lax connection for two different spectral parameters $\Lambda,\Gamma$. Following standard analysis we find
\begin{eqnarray}
& &\pb{L^A_\sigma(\Lambda),L^B_\sigma(\Gamma)}=
\frac{1}{(1-\Lambda^2)(1-\Gamma^2)}K^{AB}\partial_\sigma\delta(\sigma-\sigma')
-\nonumber \\
&-&\frac{\Gamma^2}{(1-\Gamma^2)(1-\Lambda^2)\tau_F}
L^C_\sigma(\Lambda)f_{CD}^{ \ \quad A}K^{DB}\delta(\sigma-\sigma')-\nonumber \\
&-&
\frac{\Lambda^2}{(1-\Gamma^2)(1-\Lambda^2)\tau_F}
L^C_\sigma(\Gamma)f_{CD}^{\ \quad A}K^{DB}\delta(\sigma-\sigma') \ .
\nonumber \\
\end{eqnarray}
Then, following \cite{Dorey:2006mx} we can argue that this theory possesses an infinite number of charges that are in involution and hence non-relativistic string theory on $AdS_5\times S^5$ is integrable.

\acknowledgments{This  work  was
    supported by the Grant Agency of the Czech Republic under the grant
    P201/12/G028. }

\end{document}